\documentclass[12pt,apj]{emulateapj}

\usepackage{natbib, graphics}

\usepackage{longtable}
\usepackage{amsfonts}
\usepackage{rotating}
\RequirePackage{amssymb}
\RequirePackage{latexsym}
\RequirePackage{graphicx}

\renewcommand{\deg}{\mbox{$^{\circ}$}}

\begin{document}

\title{New insights on the Galactic Bulge Initial Mass Function\altaffilmark{1}}

\shorttitle{The Galactic bulge Mass Function} 

\shortauthors{Calamida et al.}

\author{A. Calamida$^2$,
K. C. Sahu$^2$, 
S. Casertano$^2$,
J. Anderson$^2$,
S. Cassisi$^3$,
M. Gennaro$^2$,
M. Cignoni$^2$,
T. M. Brown$^2$,
N. Kains$^2$,
H. Ferguson$^2$,
M. Livio$^2$,
H. E. Bond$^2,4$,
R. Buonanno$^{3,5}$,
W. Clarkson$^6$,
I. Ferraro$^7$,
A. Pietrinferni$^3$, 
M. Salaris$^8$,
J. Valenti$^2$}

\altaffiltext{1} {Based on observations made with the NASA/ESA {\it
Hubble Space Telescope}, obtained by the Space Telescope
Science Institute. STScI is operated by the Association of Universities for
Research in Astronomy, Inc., under NASA contract NAS~5-26555.}

\altaffiltext{2}
{Space Telescope Science Institute, 
3700 San Martin Dr.,
Baltimore, MD 21218, USA;
calamida@stsci.edu}

\altaffiltext{3}
{INAF - Osservatorio Astronomico di Teramo - Via M. Maggini, sn. 64100 Teramo, Italy}

\altaffiltext{4}
{Department of Astronomy \& Astrophysics, Pennsylvania State University,
University Park, PA 16802, USA}

\altaffiltext{5}
{Dipartimento di Fisica, Universit\'a di Roma Tor Vergata, Roma, Italy}

\altaffiltext{6}
{University of Michigan-Dearborn}

\altaffiltext{7}
{INAF - Osservatorio Astronomico di Roma - Via Frascati 33, 00040 Monteporzio Catone, RM, Italy}

\altaffiltext{8}
{Astrophysics Research Institute, Liverpool John Moores University}

\begin{abstract} 
We have derived the Galactic bulge initial mass function of the SWEEPS field in the mass range
0.15 $\lesssim M/M_{\odot} \lesssim$ 1.0, using deep photometry collected with the Advanced Camera for Surveys on the Hubble
Space Telescope. Observations at several epochs, spread over 9 years, allowed us to separate the disk
and bulge stars down to very faint magnitudes, $F814W \approx$ 26 mag, with a proper-motion accuracy 
better than 0.5 mas/yr (20 km/s). This allowed us to determine the initial mass function of the pure bulge
component  uncontaminated by disk stars for this low-reddening field in the Sagittarius window. In
deriving the mass function, we took into account the presence of unresolved binaries, errors in photometry, 
distance modulus and reddening, as well as the metallicity dispersion and the uncertainties caused by
adopting different theoretical color-temperature relations. We found that the Galactic bulge initial mass function 
can be fitted with two power laws with a  break at $M \sim$ 0.56 $M_{\odot}$, the slope being steeper 
($\alpha = -2.41\pm$0.50) for the higher masses, and shallower ($\alpha = -1.25\pm$0.20) for the lower masses.
In the high-mass range, our derived mass function agrees well with the mass function derived for other regions
of the bulge. In the low-mass range however, our mass function is slightly shallower, which suggests that
separating the disk and bulge components is particularly important in the low-mass range. 
The slope of the bulge mass function is also similar to the slope of the mass function derived for the disk in the
high-mass regime, but the bulge mass function is slightly steeper in  the low-mass regime.  We used our new 
mass function to derive stellar mass--to--light values for the Galactic bulge and we obtained 2.1 $\le M/L_{F814W} \le$ 2.4 and
3.1 $\le M/L_{F606W} \le$ 3.6 according to different assumptions on the slope of  the IMF for masses larger than 1$M_{\odot}$.
 \end{abstract}


\maketitle

\section{Introduction}\label{intro}

The knowledge of the stellar initial mass function (IMF) is a fundamental piece of information in  many
research areas of astrophysics. From a theoretical point of view, providing tight constraints  on the
IMF properties in different stellar environments - both in the field and in star clusters -  is
mandatory to develop a complete and reliable theory of star formation \citep[and references
therein]{Mckee07}. At the same time, from a phenomenological point of view, the IMF is a fundamental 
property of stellar populations, and hence a crucial input in any study of galaxy formation and
evolution.  For instance, it represents an important ingredient in the computations of Population
Synthesis models  (see \citealt{vazdekis15} and references therein), and hence it affects our
capability to extract the properties of stellar  populations such as their luminosity evolution over
time, the mass--to--light ratio, the total star formation  rate at low and high redshifts, and so on.
Therefore, it appears evident that to improve our knowledge of the IMF, or at least to have stronger
observational constraints on this crucial ingredient, is of pivotal importance in many astrophysics
research fields.

It is particularly important to analyze the properties of the IMF in various stellar environments
such as the disk and the bulge of spiral galaxies in order to verify whether the well-known differences
(in age and chemical composition) in the stellar populations hosted by the distinct galactic components
have an impact on the IMF \citep{zoccali03}. An additional reason that makes the study of the IMF in 
the bulge of spiral galaxies and elliptical galaxies important is due to the possibility that these
spheroids  could potentially contain the majority of the stellar mass of the universe (see, for
instance \citet{fukugita98}. 

As we discuss, the IMF for the Galactic bulge is unlikely to be very different 
from the present-day mass function (PDMF) below the main-sequence turn-off (MSTO), 
since most of the star-formation in the Galactic bulge happened within about 2 Gyr \citep{clarkson08}, 
with no evidence of star formation after that. So we will refer to the observed PDMF of the Galactic bulge as the IMF 
in the mass range below the MSTO, which occurs at $\approx$1.0 $M_{\odot}$ for a stellar population with 
solar metallicity and an age of $t$ = 11 Gyr  \citep[hereafter Paper I]{calamida14b}.

In spite of the huge improvements achieved in the observational facilities, there is not yet any chance
to directly measure the IMF of spheroids outside of our Galaxy. As a consequence, the measurement  of
the IMF in the Galactic bulge is a fundamental benchmark (or reference point) for any analysis devoted
to investigate this property in extra-galactic spheroids \citep{calchi08}. 

Concerning the Galactic bulge, the two most recent determinations of the IMF in our spheroid have  been
performed by \citet[hereinafter HO98]{holtz98} and by \citet[hereafter ZO00]{zoccali00}, by taking
advantage of the exquisite observational capabilities of the  {\sl Hubble Space Telescope} ({\it HST}).
In particular, the analysis performed by ZO00 pushed a step forward the knowledge of  the bulge IMF
thanks to the use of the Near-Infrared Camera and Multi-Object Spectrometer (NICMOS) available at that
time:  the derived mass function still represents the deepest measured to date and extends to $\sim$
0.15 $M_{\odot}$.  They found a power-law slope for the IMF equal to $\alpha=$ $-1.33$ (when a Salpeter
IMF would have $\alpha = -2.35$, where $dN/dM$ = $C \times M^{\alpha}$.), with some hint for a
possible change of the power slope - $\alpha \approx$ $-2$ at $\sim$ 0.5 $M_{\odot}$.  ZO00 also found
that the derived bulge IMF is steeper than that measured for the Galactic disk \citep{reid97, gould97}.
In this context, it is also worth noting that \citet{dutton13} have recently used strong lensing and
gas  kinematics to investigate the existence of possible differences in the properties of the IMF
between the disk and the bulge in a sample of  spiral galaxies within the Sloan WFC Edge-on Late-type
Lens Survey \citep[SWELLS]{treu11}. As a result they found a significant  difference between the bulge
IMF and that of the disk, the former being more consistent with a Salpeter IMF, and the latter being
more consistent with a Chabrier-like IMF.

On the basis of this evidence, it appears quite appropriate to analyze the properties of the  Galactic
bulge IMF in different fields of view and using more updated observational datasets. In a previous
paper  \citep[hereinafter Paper I]{calamida14b}, we have taken advantage of the availability of a huge
photometric  dataset for the low-reddening Sagittarius window in the Galactic bulge collected with the
{\sl Advanced Camera for Survey}  (ACS) on board {\it HST}, to obtain the first unambiguous detection
of the white dwarf cooling sequence of the  Galactic bulge. In this manuscript, we use the same data to
perform a  thorough analysis of the IMF in this field of the bulge in order to provide additional, 
independent insights on the bulge IMF,  thus supplementing the results of previous analyses. In this
investigation we explore a larger and denser field compared to what was previously observed by HO98  in
the Baade's Window and by ZO00 in a field at $(l = 0.277^\circ,,b = -6.167^\circ,)$. Most importantly, for the first
time we estimate the Galactic bulge IMF based on a clean sample of bulge stars  thanks to the very
accurate proper motions  (down to $F814W \approx$ 26 mag) that we were able to measure.  Furthermore,
we use a statistical approach to apply a correction for the presence of unresolved binaries.
We note that the slope of the very low-mass range of the IMF is fundamental to estimate the mass budget of 
a stellar population, since a major fraction of the stellar mass is included in this range and 
low-mass stars have been hypothesized to contain a significant fraction of the total 
mass in the universe \citep{fukugita98}.
It is even more important in the case of the Galactic bulge since this component might 
include $\approx$20\% ($1.8\times10^{10} M_{\odot}$) of the mass of the Galaxy \citep{sofue09, portail15}.

The structure of the paper is as follows: in \S 2 we discuss the observations and data reduction in
detail,  while in \S 3 we describe how we selected a clean sample of bulge stars.  In \S 4 we present
the theoretical mass--luminosity relations we adopted to convert the luminosity functions in mass
functions, while \S5 deals with the different systematics that affect the estimate of the initial mass
function. In \S6 we compare the derived IMF for the bulge with the disk mass function, and in \S 7 we
derive a minimum value for the stellar mass--to--light ratio of the Galactic bulge. \S 8 deals with the
gravitation microlensing events predicted by the bulge IMF derived in this work, and the conclusions
are presented in \S 9.

\section{Observations and data reduction}\label{obs}

We observed the Sagittarius Window Eclipsing Extrasolar Planet Search (SWEEPS) field ($l = 1.25^\circ, b =
-2\fdg65$) in the Galactic bulge in 2004 and again in 2011, 2012 and 2013 with {\it HST}, using the
Wide-Field Channel of ACS (proposals GO-9750, GO-12586, GO-13057, PI: Sahu). The SWEEPS field covers
$\approx 3\farcm3 \times 3\farcm3$ in a region of relatively low extinction in the bulge ($E(B-V)
\lesssim 0.6$~mag; \citealt{oosterhoff}). The 2004 observations were taken in the $F606W$ (wide $V$)
and $F814W$ (wide $I$) filters over the course of one week (for more details see \citealt{sahu06}). The
new data were collected between October 2011 and October 2013, with a $\sim$ 2-week cadence, for a
total of  60 $F606W$- and 61 $F814W$-band images.  The two datasets, the 2004 and the 2011--2012--2013
(hereafter 2011--13),  were reduced separately by using a software that performs simultaneous
point-spread function (PSF)  photometry on all the images. The choice to reduce the two datasets
separately is due to the high relative proper motions of the disk and bulge stars in this field, caused by the Galactic
rotation: the disk star relative proper motions (PMs) peak at $\mu_l \approx$ 4 mas/yr, with a
dispersion of $\approx$ 3 mas/yr, whereas the  bulge motions are centered at  $\mu_l \approx$ 0 mas/yr, with a
dispersion of $\approx$ 3 mas/yr (see Paper I).  This means that a substantial fraction ($\sim$ 30\%)
of stars would move by more than half a  pixel (25 mas) in 9 years. 

We calibrated the instrumental photometry to the Vegamag system by adopting the 
2004 photometric zero-points, and we obtained a catalog of $\approx$ 340,000 stars
for the 2004 and for the 2011--2013 datasets.
The left panel of  Fig.~1 shows the 
$F814W,\, (F606W - F814W)$ Color-Magnitude Diagrams (CMD)
for all the observed MS stars in the 2011--2013 dataset. 

The right panel shows 
the sample completeness as a function of the $F814W$ magnitude.
Details on how the completeness was derived are given in \S \ref{art}.
This figure shows that the completeness is $\sim$ 50\% at
$F814W \sim$ 25.5 mag. The completeness of the $F606W$ magnitude is  $\sim$ 50\% at $F606W \sim$ 28 mag.
The  2004 dataset has a very similar completeness, reaching $\sim$ 50\% at
$F814W \sim$ 25.3 mag and $F606W \sim$ 28 mag, respectively.

In order to obtain a clean bulge MS sample to derive the IMF, we estimated 
the PMs of the stars in this field by combining the astrometry and the photometry of the 2004 and the 2011-2013 datasets.
By comparing the positions of stars in the two epochs we estimated PMs for $\approx$ 200,000 stars 
down to $F814W \approx$ 25.5 mag. 

\begin{figure*}
\begin{center}
\label{fig1}
\includegraphics[height=0.7\textheight,width=0.57\textwidth, angle=90]{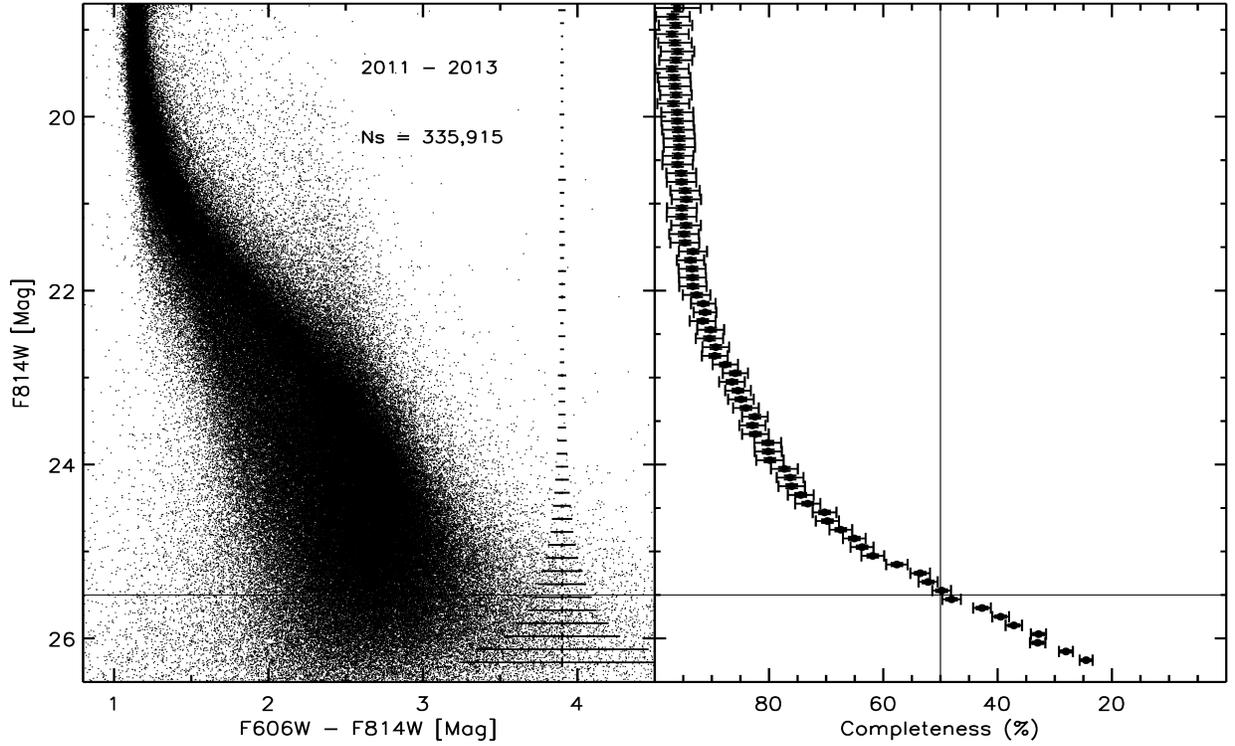} 
\caption{Left: $F814W,\ (F606W - F814W)$ CMD of MS stars in the SWEEPS field based on the 2011--2013 dataset.
Error bars are also labelled. Right: completeness of MS stars as a function of the $F814W$ magnitude. 
The horizontal lines in both panels represent the $F814W$ magnitude at which the completeness is 50\% and the 
vertical line in the right panel represents the 50\% completeness level.}
\end{center}
\end{figure*}

\subsection{Artificial star tests}\label{art}

To properly characterize the completeness of the measured PMs, the photometric errors and the  
errors due to the reduction and selection techniques adopted, we performed several artificial star (AS) tests.
We created a catalog of $\approx$ 200,000 artificial MS stars, with magnitudes
and colors estimated by adopting a ridge line following the MS.
We then produced a second artificial star catalog, by using the same input magnitudes and colors,
but applying a PM to each star. We assumed the bulge PM distribution as measured in Paper I, 
with the distribution centered at $\mu_l \approx$ 0 mas/yr, and a dispersion of $\approx$ 3 mas/yr. 
Artificial stars were added and recovered one by one on every image of the two datasets
by using the same reduction procedures adopted earlier. In this way the level of 
crowding is not affected.
In order to estimate the magnitude and color dispersion of the MS due to photometric errors and
data reduction systematics, we selected recovered artificial
stars with $\Delta Mag = (Mag_i - Mag_o) \le$ 0.5 mag, and $d = \sqrt{(X_{o}-X_{i})^{2}+(Y_{o}-Y_{i})^{2}} \le$ 0.75 pixel, 
where the quantities with subscript $i$ represent the input, and $o$ represent the output, 
in both datasets, ending up with a sample of 146,225 stars. 
We applied this selection because stars which were not recovered in a circle of radius 0.75 pixel can be safely 
considered not found.
The left panel of Fig.~2 shows the selected recovered artificial stars for the 2011--2013 dataset (red dots) 
plotted in the $F814W, (F606W - F814W)$ CMD; the observed stars are plotted as well as grey dots. 
The right panel shows the recovered color spread of the MS as a function of the $F814W$ magnitude.  
The comparison of the artificial (red dots) and observed (grey dots) CMDs indicates that we
are not able to reproduce the entire color spread of the MS by assuming the presence of 
a single stellar population of solar metallicity and age $t$ = 11 Gyr.  
A metallicity spread of more than 1 dex is present in the SWEEPS 
bulge field based on medium-resolution spectra of MS turn-off, sub-giant and red-giant branch 
stars collected with FLAMES at the Very Large Telescope (see Paper I for more details).
The metallicty spread can further broaden the MS, and differential reddening, 
depth effects as well as binaries might also play a role.
It is worth mentioning that most stars in the color and magnitude ranges 
2.0$ < F606W - F814W <$ 2.5 and 18.5 $< F814W < $21.5 mag belong to the (closer) disk population.

\begin{figure*}
\begin{center}
\label{fig2}
\includegraphics[height=0.7\textheight,width=0.57\textwidth, angle=90]{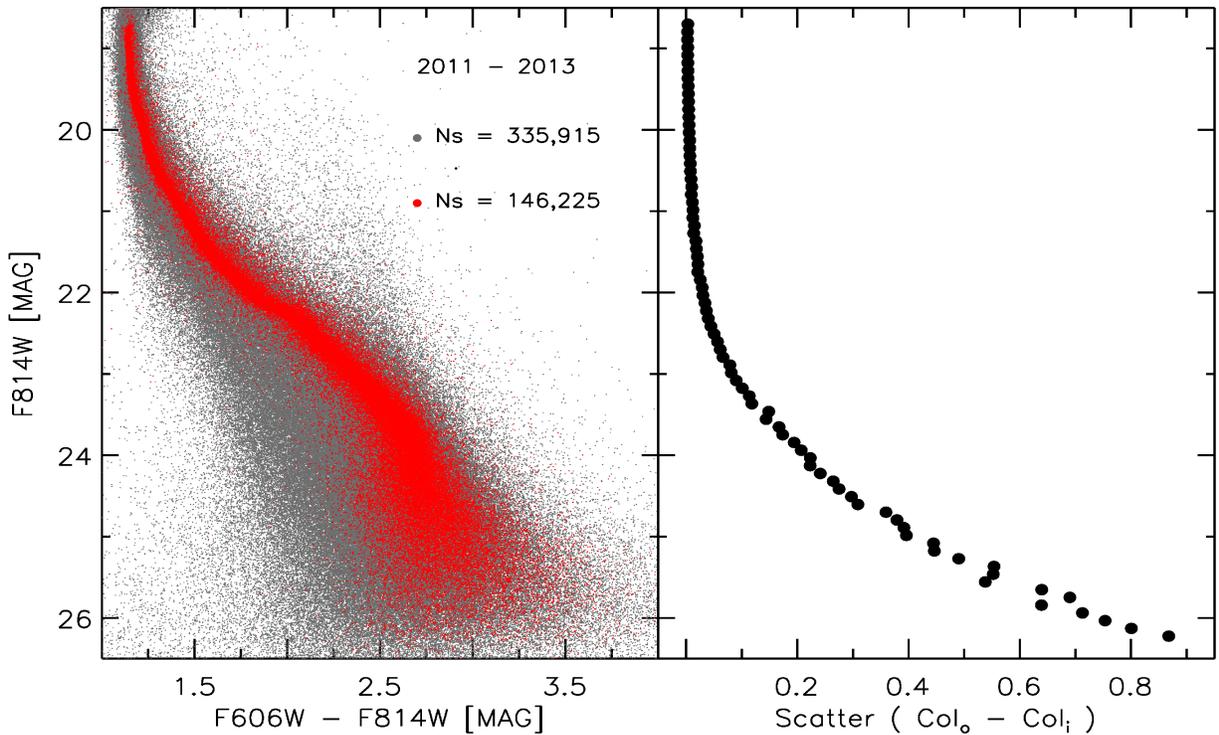} 
\caption{Left: $F814W,\ (F606W - F814W)$ CMD of recovered artificial stars for the 2011--2013 dataset (red points).
Observed stars are marked with grey points.
Stars are selected in magnitude, $\Delta Mag \le$ 0.5 mag, and in position, $\Delta d \le$ 0.75 pixel.
Right: ($F606W - F814W$) photometric scatter as a function of the $F814W$-band magnitude 
as estimated from the artificial star test.}
\end{center}
\end{figure*}

In order to estimate the completeness of the measured PMs, we matched the two recovered sets 
of artificial stars and compared the output with the input PMs in the direction of both $X$ and $Y$ axes
as a function of the two magnitudes. 
Fig.~3 shows this comparison for the X (top panel) and the Y (bottom) axes versus the $F814W$ magnitude.
Only stars with $\Delta Mag \le$ 0.5 mag and $d \le$ 0.75 pixel are shown. 
This plot shows that the dispersion of the recovered PMs increases at fainter magnitudes as expected and
 the accuracy of the measured PMs is better than 0.1 mas/yr  ($\approx$ 4 km/s at the distance of the Galactic
bulge) at magnitudes brighter than $F814W \le$ 23.
At $F814W \sim$ 25 mag where the completeness is $\gtrsim$ 50\% for both datasets, 
the recovered PM scatter is $\approx$ 0.25 mas/yr ($\approx$ 10 km/s) within 3 $\sigma$ uncertainties. 
This precision will allow us to separate bulge from disk stars down to very faint magnitudes and to characterize the 
Galactic bulge mass function down to the very low-mass (VLM) range, $M <$ 0.5 $M_{\odot}$.

\begin{figure}
\begin{center}
\label{fig3}
\includegraphics[height=0.65\textheight,width=0.5\textwidth]{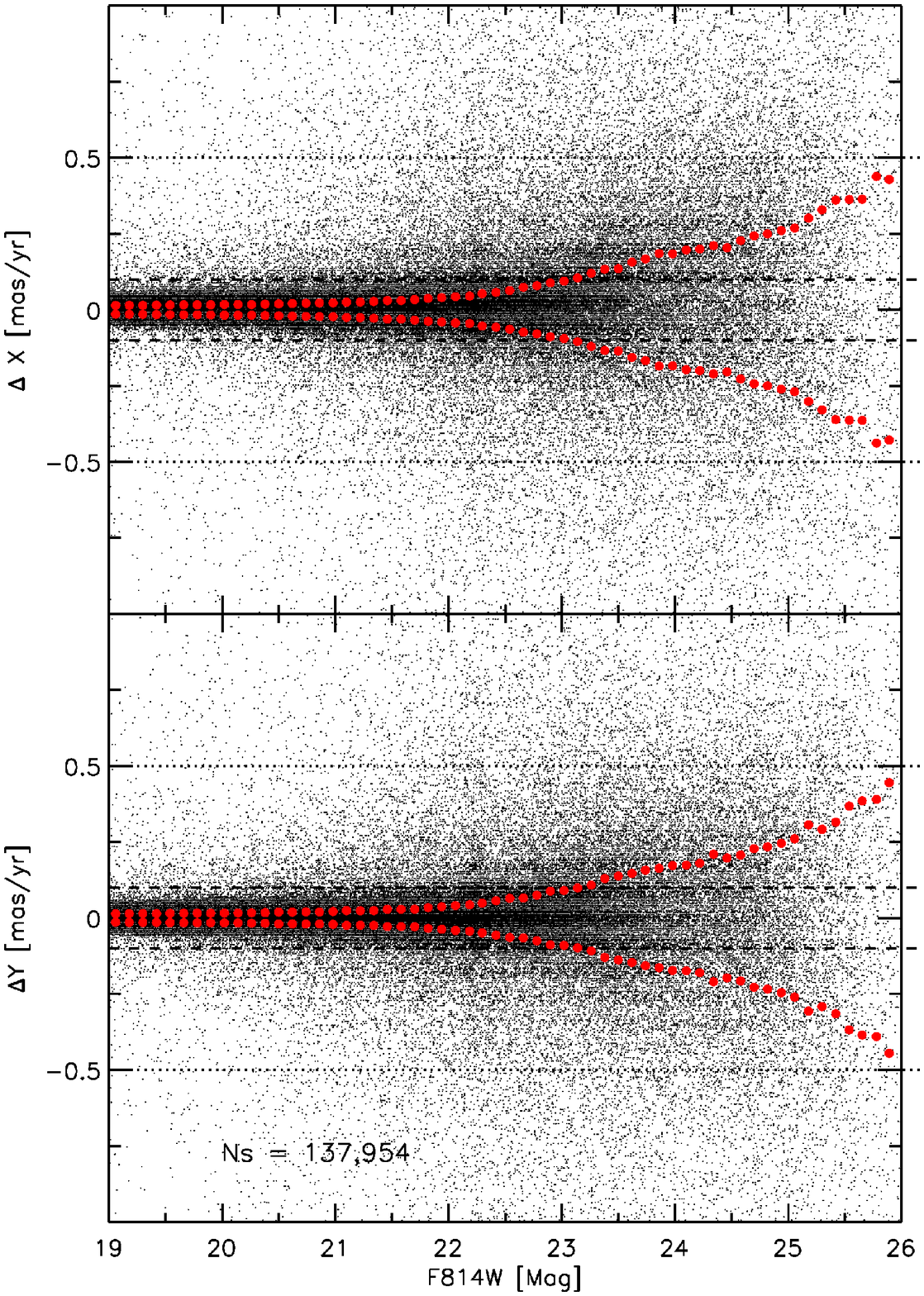} 
\caption{Comparison of Input and Output proper motions in the X (top panel) and Y (bottom) 
axes of stars recovered in the AS test as a function of the $F814W$ magnitude.
The 3$\sigma$ limit is indicated by the overplotted red dots. Dashed and black dotted 
lines mark a dispersion of 0.1 and 0.5 mas/yr, respectively.
}
\end{center}
\end{figure}

\section{A clean bulge main-sequence sample}\label{bulge}

We adopted the measured PMs to select a sample of main-sequence (MS) stars devoid of disk-star contamination from 
the 2011--2013 dataset. 
PMs are projected along the Galactic coordinates and we considered stars with 
$\mu_l \le -2 \,\rm mas\,yr^{-1}$ to belong to the bulge, following the criteria adopted in Paper I. 
This selection allowed us to keep $\approx$ 30\% of bulge members while the residual contamination of the sample 
by disk stars is $\lesssim$ 1\%. We ended up with a sample of 67,765 bulge MS stars. 
Note that the total contamination by disk stars in the SWEEPS field is $\approx$ 10\%, as shown 
in the previous work of \citet{clarkson08}.
Fig.~4 shows the $F606W$ (left panel) and the $F814W$ (right) versus $(F814W-F606W)$ CMDs of the selected bulge MS stars.
The magnitude range covered by MS stars is different when observing with the $F606W$ or the $F814W$ filter, 
decreasing from $\sim$ 8.5 to 7 magnitudes. This happens because very low-mass MS stars are 
cooler and thus more luminous at longer wavelengths. 

The CMDs of Fig.~4 show that the color spread of the MS did not substantially decrease compared to the CMD of 
Fig.~1, confirming that most of the color dispersion is due to the spread in metallicity, the presence of some amount 
of differential reddening, depth effects and binaries. 

Fig.~5 shows the observed PM-cleaned bulge MS luminosity function (dashed line) based on the $F814W$ magnitude 
for the stars plotted in the CMDs of Fig.~4. The completeness measured from the AS test is used to 
correct the number of observed stars per magnitude bin and the corrected luminosity function is 
over-plotted in the same figure with a solid line.
We applied the completeness correction by binning on the observed magnitudes; in this way we take into
account the uncertainties due to the photometric errors moving the stars among the magnitude bins.
The $F814W$-band corrected luminosity function of Fig.~5 extends from just below the bulge 
MSTO at $F814W \sim$ 19 mag down to $F814W \sim$ 26 mag, where the completeness level is $\sim$ 30\%. 
A similar luminosity function is obtained by adopting the $F606W$ magnitude.

\section{The mass--luminosity relation}\label{mass}

In Paper I we used the BaSTI \footnote{http://albione.oa-teramo.inaf.it/} \citep{pietrinferni04, pietrinferni06} 
stellar-evolution database to fit isochrones to the bulge CMD. 
In order to extend the BaSTI isochrones\footnote{In their standard format the minimum initial mass 
in the BaSTI isochrones is equal to ${\rm 0.5 M_\odot}$. 
The BaSTI isochrones extended in the VLM star regime are available at the BaSTI URL repository.} 
to the range of very-low-mass stars (${\rm M < 0.5M_\odot}$) we computed very-low-mass (VLM) 
stellar models for exactly the same chemical composition of the BaSTI ones, by adopting the same physical 
inputs used in \citet[hereafter CA00]{cassisi00}. 
We note that, the accuracy and reliability of the BaSTI models and 
their extension to the VLM stellar regime have been extensively tested by comparing 
them with observed CMDs and mass-luminosity (M-L) datasets for both field and 
cluster stars. As a result, a very good level of agreement has been obtained with 
the various observational constraints \citep{cassisi00, bedin09, cassisi09, 
cassisi11, cassisi14}.
Since the VLM stellar models have been computed by using a different physical framework
compared to the models of more massive stars in the BaSTI library (see \citet{pietrinferni04} 
and CA00 for more details on this issue) one can expect that, 
in the stellar mass regime corresponding to the transition between the BaSTI and the 
VLM stellar models occurring at about ${\rm \sim 0.6M_\odot}$, some small mismatch 
in surface luminosity and effective temperature at a given mass is possible. 
Since in retrieving the IMF one has to rely on the first derivative of the theoretical 
M-L relation, it is important to eliminate any such discontinuity in the M-L relation \citep{kroupa97}. 
To this aim, we devoted a huge effort - which included computing additional stellar models 
using both the physical inputs adopted for the BaSTI library and that used by CA00 - in order to 
match the two model datasets at the stellar mass with (almost) the same luminosity
and effective temperature.

The evolutionary predictions were transformed from the theoretical to the observational plane by adopting 
the color--$T_{\rm eff}$ relations and bolometric correction scale for the ACS filters provided by \citet{hauschildt99} 
for $T_{\rm eff}  \le 10,000$~K, while at larger $T_{\rm eff}$ we adopted the relations published by \citet{bedin05a}. 

Fig.~6 shows selected scaled-solar isochrones\footnote{Our referee correctly pointed out that bulge stars 
appear to be $\alpha$-enhanced up to about solar metallicity \citep{zoccali08, gonzalez11, johnson11, johnson13}. 
However, we decided in present work to adopt scaled-solar
models due to the lack of suitable alpha-enhanced VLM star sequences in a wide metallicity range.
This notwithstanding, we note that all the comparisons performed in present paper are performed 
at constant global metallicity (and not at constant $[Fe/H]$ and it is well known that $\alpha$-enhanced 
stellar models are nicely mimicked by scaled-solar one with the same global metallicity (see, e.g.
Pietrinferni et al. 2006 and references therein).} for the same age, $t$ = 11 Gyr, and different metallicities,
$Z$= 0.008, 0.0198, 0.03, plotted in the $F814W$ versus $log (M/M_{\odot})$ plane. 
We selected models with this age and abundances based on the fit of the bulge CMD performed in Paper I (see Fig.~2)
and on the spectroscopic metallicity distribution for this field.
In the same plot a solar metallicity isochrone but for an age of 8 Gyr is also shown (blue solid line). 
As expected, in the explored age and stellar mass range, the M-L is completely unaffected by an age change. 
In order to check the impact on the adopted M-L relation related to the use of a different bolometric 
correction scale, we also plotted the 11 Gyr, solar metallicity isochrone transferred in the observational plane by using 
the standard Johnson bolometric corrections provided by \citet{pietrinferni04} and the transformations 
from the Johnson to the HST photometric system by \citet[red solid]{sirianni05}.

The five mass--luminosity relations all show a slight change of the slope around 
$log (M/M_{\odot}) \approx$ -0.3 ($M \approx$ 0.5 $M_{\odot}$).
This inflection point is due to the molecular Hydrogen recombination occurring at a mass 
equal to $\approx 0.5M_\odot$; the formation of the $H_2$ molecule changes the value of the 
adiabatic gradient and, hence, the stellar structure thermal stratification (see Cassisi et al. 2000 and references therein).

Fig.~6 also shows the impact of using various metallicities or ages for the selected M-L relation. 
As discussed, for old ages, $t \ge 8$ Gyr, suitable for the Galactic bulge population under scrutiny, 
the exact value of the selected age is quite irrelevant. 
On the other hand, the change in the mass derived (at a fixed magnitude) using two 
different mass-luminosity relations corresponding to $Z=0.008$ (which is the most 
metal-poor chemical composition we selected) and $Z=0.03$ (which is our most 
metal-rich composition) is only about $\approx$ 0.04--0.08 $M_{\odot}$ in the high-mass range ($M >$ 0.5  $M_{\odot}$), 
and $\approx$ 0.02--0.04 $M_{\odot}$ in the lower-mass range.
The spectroscopic metallicity distribution we derived for the SWEEPS field, as discussed in \S 3.1 and Paper I, 
spans a range of metallicity from $[M/H] \sim$ -0.8 to $\sim$0.6,  i.e. more than 1 dex. However, the 
distribution shows three peaks at $[M/H] \sim$ -0.4, 0.0 and $0.3$ and most of the stars, $\sim$ 85\%, 
are included in the range -0.5 $< [M/H] <$ 0.5. We can then safely assume the aforementioned metallicity values, 
$Z = 0.008 and Z = 0.03$, corresponding to the more metal-poor and the more metal-rich peaks of the distribution,
to test the effect of metallicity on the mass estimate.
However, we also tested the effect of further decreasing the metallicity of the
adopted models, by using an isochrone for $Z = 0.002$ and the same age, $t =$ 11 Gyr, to convert luminosities into masses.
In this case, the mass estimate changes by $\sim$ 17\% in the entire mass range, 
when going from the more metal-rich model, $Z = 0.03$, to the more metal-poor, $Z = 0.002$.
For a small fraction of stars in our field, less than $\sim$ 10\%, the mass estimate will have a $\sim$ 5\% larger uncertainty.

The impact of using a different bolometric correction scale for transferring the models from the theoretical 
to the observational plane in the derived masses is smaller and of the order of $\approx$ 0.005 $M_{\odot}$ 
in the entire mass regime.

\begin{figure*}
\begin{center}
\label{fig4}
\includegraphics[height=0.7\textheight,width=0.57\textwidth, angle=90]{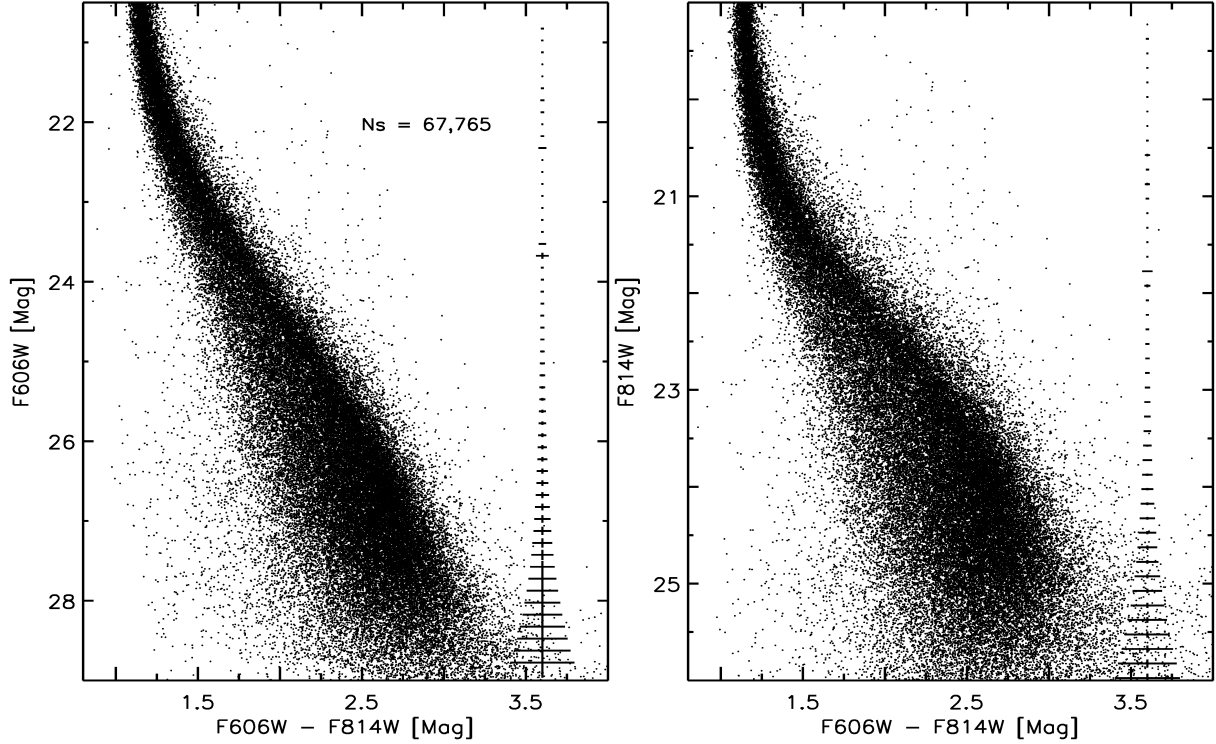} 
\caption{Left: PM-cleaned bulge MS $F606W,\ F606W - F814W$ CMD; note that 70\% of the bulge stars 
were rejected because of the PM selection. Error bars are also labelled.
Right: Same stars plotted on the $F814W,\ F606W - F814W$ CMD.}
\end{center}
\end{figure*}

\section{The Galactic bulge initial mass function}\label{initial}

The mass-luminosity relation we obtained for MS stars by using 
BaSTI isochrones is only the first step towards determining the IMF of
the Galactic bulge. Uncertainties due to the assumed distance 
and reddening, presence of differential reddening, 
metallicity dispersion, depth effects, and the presence of binaries need to be
taken into account.

Following \citet{sahu06} and Paper I, we fitted the bulge CMD using a distance modulus of $\mu_0 = 14.45$ mag \citep{sahu06}
and  a mean reddening of $E(B-V)$ = 0.5 mag and a standard reddening law.
Extinction coefficients for the WFC filters are estimated by applying the \citet{cardelli89} reddening relations 
and by adopting a standard reddening law, $R_V = A_V/E(B-V) = 3.1$, finding $A_{F606W} = 0.922 \times A_V$,  
$A_{F814W} = 0.55 \times A_V$, and $E(F606W - F814W)= 1.14 \times E(B-V)$.
It is worth mentioning that if we use the reddening value estimated by \citet{nataf13} for the SWEEPS field, 
$E(V-I) = 0.79$, and their extinction curve, $R_V = A_V/E(B-V) = 2.5$, we obtain $E(B-V) = 0.47$, 
in good agreement with the value we assumed.

We used the $F814W$-band luminosity function to probe 
the bulge IMF since MS stars are brighter at redder colors and so the photometry in this filter is
more complete and accurate than in the $F606W$ filter for the same mass (see the CMDs in Fig.~4).
We converted observed magnitudes into masses using the mass-luminosity relation for solar metallicity, $Z$ = 0.0198, and for
an age of $t$ = 11 Gyr, transformed by using the color--$T_{\rm eff}$ relations by \citet{hauschildt99}.
As we showed in the previous section, age does not significantly affect the mass-luminosity relation for $t \ge$ 8
Gyr, and observational evidence shows that most bulge stars in our field are older than 8 Gyr \citep[Paper I]{clarkson08}.

\begin{figure}
\begin{center}
\label{fig5}
\includegraphics[height=0.4\textheight,width=0.5\textwidth]{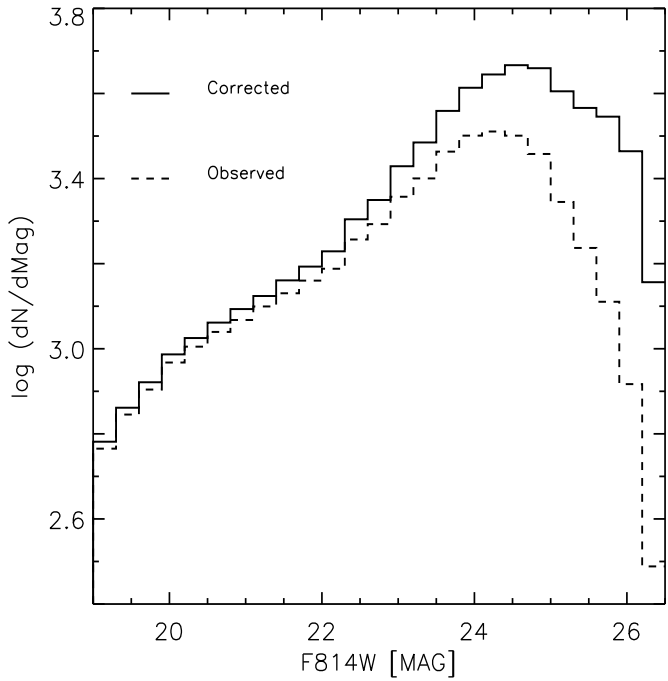} 
\caption{$F814W$-band observed luminosity function for the Galactic PM-cleaned bulge MS stars  (dashed line) 
and the luminosity function corrected for completeness (solid line). }
\end{center}
\end{figure}

In order to estimate the effect of dispersion in metallicity, we computed the difference in the masses 
derived by adopting three different metallicities: solar ($Z = 0.00198$), metal-rich ($Z =
0.03$), and metal-poor ($Z = 0.008$).  For magnitudes in the range $18.0 < F814W < 26 $, this changes 
the inferred masses by 0.02 to 0.08 $M_{\odot}$, resulting in an uncertainty of $\approx$ 8\% in mass.

We also varied the assumed distance modulus by 0.2 mag, from 14.35 to 14.55 mag, 
corresponding to a depth of $ \sim$ 1 Kpc, and the extinction, $E(B-V)$, from 0.45 up to 0.55 mag. 
Both the distance and reddening uncertainty affects the derived masses by $\approx$ 0.01
$M_{\odot}$ over the entire mass range, i.e. 1--5\%.
Similarly, adopting different color--$T_{\rm eff}$ relations changes the derived massed by less than 2\%.

By summing in quadrature the uncertainties related to the parameters of the bulge CMD as fitted 
to our data, including metallicity, distance, reddening, and color--$T_{\rm eff}$ relations, 
we obtain a final systematic uncertainty on the mass estimate for each star.
This uncertainty varies depending on the inferred mass and is carried through the remainder of the analysis.

\begin{figure}
\begin{center}
\label{fig6}
\includegraphics[height=0.4\textheight,width=0.5\textwidth]{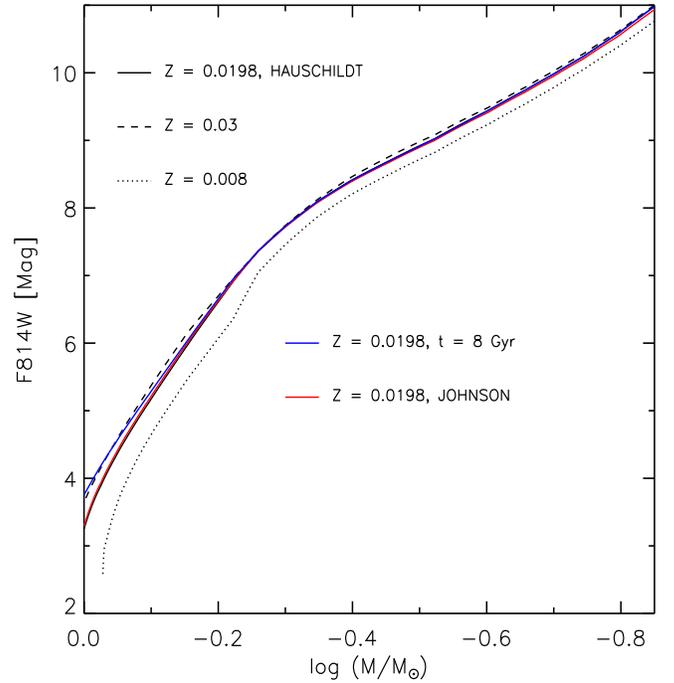} 
\caption{Theoretical mass-luminosity relations for different metallicities, ages, and color-temperature relations.}
\end{center}
\end{figure}

\subsection{The effect of unresolved binaries}

Unresolved binaries, i.e. the expected presence of equal or lower-mass binary companions 
for many of the main sequence stars we observe, are likely to affect the inferred IMF 
of the bulge especially at lower masses, such as $M < 0.5 M_\odot $ (see, e.g., \citet{kroupa91} and \citet{kroupa01}).

The availability of photometry in two different filters for a large fraction of our stars potentially allows us to correct 
for the effect of unresolved binaries, as binary systems will be somewhat redder than single stars
of the same apparent brightness. However, the photometry, in particular at the faint end, 
is not sufficiently accurate for a direct identification of 
individual binary systems; our correction must therefore be probabilistic.

Both the fraction of binaries and the distribution of mass ratios for the Galactic bulge are not well constrained.  
However, in Paper I we showed that there is evidence for a substantial fraction of He-core 
white dwarfs in the bulge based on the color dispersion of the cooling sequence and the comparison
between star counts and predicted evolutionary lifetimes. 
 According to standard stellar evolution models, He-core white dwarfs can only be produced in
a Hubble time by stars experiencing extreme mass-loss events, such as in compact binaries.  Indeed, in Paper I we 
reported our finding of two dwarf novae in outburst and five candidate cataclysmic variables in the same field, 
both of which are characteristic of a population of binaries.  Our evidence at the time suggested that the 
Galactic bulge has a fraction of binaries of larger than 30\%.

For the present analysis, we assume that the distribution of mass ratios of binary stars in 
the bulge follows the distribution derived by \citet[hereafter DM]{duquennoy91} for a sample 
of 164 F- and G-dwarf  stars in the solar neighborhood. 
The distribution is a log-normal and it is described by the functional form:
\begin{equation}
\xi (q) = C e^{\left \{ \frac{-(q-\mu)^{2}}{2 \sigma_q ^{2}} \right \}}
\end{equation}
in the interval [0,1], where $q = M_2/M_1$, $\mu$ = 0.23 and $\sigma_q$ = 0.42 and $C$ = 10,900 for our sample of bulge stars.

We also repeated the experiment by assuming a flat mass-ratio distribution similar to the
distribution found by \citet{raghavan10} based on data for 454 F- to K-dwarf stars within 25 pc of the Sun.

\begin{figure}
\begin{center}
\label{fig7}
\includegraphics[height=0.4\textheight,width=0.5\textwidth]{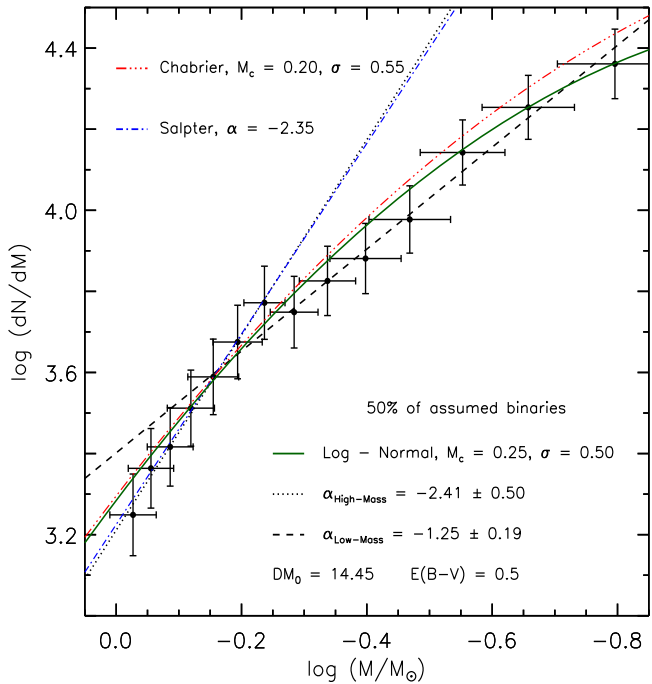}  %{IMF.ps}  
\caption{IMF of the Galactic bulge. The two power laws that fit the IMF are over plotted,
for a slope $\alpha$ = -2.41$\pm$0.50 (dotted line) and $\alpha$ = 1.25$\pm$0.19 (dashed),
together with a log-normal function with $M_c =$ 0.25 and $\sigma$ = 0.50 (green solid). 
The Salpeter mass function (blue dashed-dotted line) and the Chabrier log-normal
function (red dashed double dotted) are also shown.
Error bars are displayed. }
\end{center}
\end{figure}

In a simplified Bayesian approach, we use the fraction of binaries and the distribution of mass 
ratios from Equation (1) as a prior for the presence and mass of binary companions, 
and then use the observed photometry to determine the posterior probability 
distribution of companion mass for each star in our sample.

For simplicity, for each observed bulge star we consider 11 discrete options $J, J = 0, \dots, 10 $. 
The option $ J = 0 $ implies a single star, $ J > 0 $ implies a binary system with mass ratio $ q_J =  J / 10 $.
The prior probability $ Pr_J $ of each option is consistent with the DM distribution 
with an overall binary fraction of 50\%; thus 
$ Pr_0 =$  0.5, $Pr_{1-10} = $ 0.07, 0.072, 0.07, 0.063, 0.06, 0.05, 0.04, 0.035, 0.025, 0.015.

For each value of $ J $, the total system mass $ (M_T)_J = (M_1)_J + (M_2)_J $ is chosen 
so as to match the total flux in the $F814W$ filter, using the appropriate mass-luminosity
relation for age, metallicity and 
distance for both components (or only one component if $ J = 0 $).  
We then compute the likelihood $ P_{D|M} = P (\hbox{Data} | \hbox{Model}) $ of the measured total
flux in the $F814W$ filter, given the model, using a Gaussian distribution for the
flux with the realistic photometric errors derived above.  
To the photometric errors derived from the AS tests,
we added errors due to the presence of a metallicity spread, differential reddening
and depth effects. These have been derived by using mass-luminosity 
relations for different metallicities, and by varying the distance modulus of 0.2 mag, and 
reddening of 0.1 mag, as described in \S 5.
To each observed star we thus assign a probability distribution function (PDF) of the 
component masses over the allowed values of the mass ratio between components, according to
the classic Bayes formula: $ P_J = Pr_J * P_{D|M} / P (\hbox{Data}) $,
where $ P(\hbox{Data}) $ is a normalization factor chosen to take into account the 
estimated completeness correction.  

Finally, we generate multiple realizations of the full stellar mass function by 
randomly drawing stellar distributions with the probabilities thus determined. This 
procedure allows us to better understand and quantify the uncertainties arising from 
the correlated nature of the probabilities for each object (e.g., only one value of the mass 
ratio can be selected for each system).

\begin{figure}
\begin{center}
\label{fig8}
\includegraphics[height=0.37\textheight,width=0.5\textwidth]{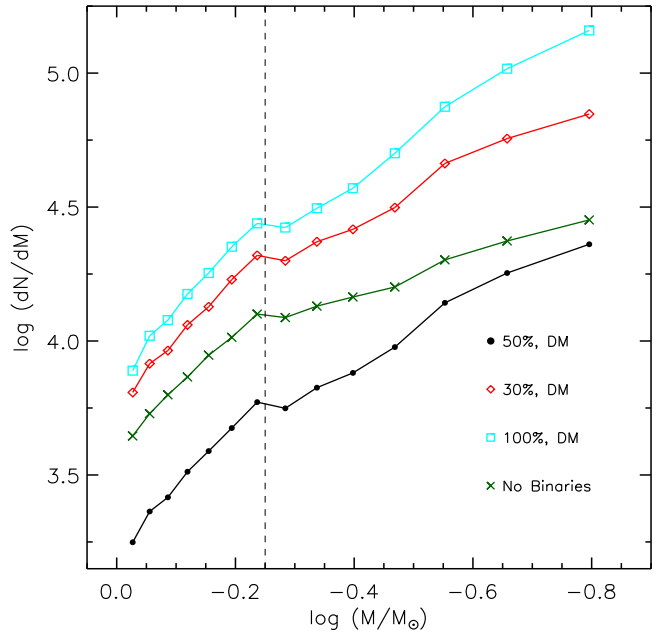}  
\caption{IMFs of the Galactic bulge derived by assuming different fraction of binaries and
a DM mass-ratio distribution for the binaries.
 }
\end{center}
\end{figure}

In practice, larger mass ratios $ q_J $ generally correspond to redder 
$F606W - F814W$ colors at given $F814W$-band flux; thus stars that lie red-ward
of the main sequence of Fig.~4  will generally favor larger mass
ratios, while stars located near the main sequence will be consistent with a single 
star or a low-mass binary companion which contributes little to the total flux. 
However, note that for many stars the photometric error is large enough 
that photometry (through the term $ P_{D|M} $) does not provide a strong discriminant; for such cases,
the final probability $ P_J $ for each option is the same as the prior 
probability $Pr_J$. By not taking into account the photometric color
information, for instance, the distribution changes by $\approx$ 2-7 \% in the VLM range, 
and by less than 1\% at higher masses, i.e. M $>$ 0.5 $M_{\odot}$.

As discussed in the following subsection, undetected binaries have a substantial
impact in the inferred mass function, especially below $\approx$ 0.5 $M_\odot$.
However, we must remark that the distribution of binary properties is uncertain
and poorly constrained by the data at hand; changing the assumed binary 
fraction and the a priori distribution of mass ratios would also alter
the derived mass function, as we show in Section 5.2.

The treatment above is somewhat simplified in comparison with a fully Bayesian 
approach, in which we would consider fully the uncertainties in 
in the parameters of the model (metallicity, distance, reddening variation), using for 
each an appropriate distribution rather than the ``best" values. 
We defer this more complex and computationally expensive approach to the analysis 
of the full data set, including one more season of photometry and eleven additional fields.

\subsection{Discussion}\label{discuss}

One of the realizations of the Galactic bulge IMF is shown in Fig.~7. 
Error bars also include the uncertainties that come from statistical noise in the star 
counts.
We generated 10,000 realizations of the same mass function and fitted them by adopting two 
power laws. The fit was performed by varying the mass break-point in the range 
$-0.2 \le log (M/M_{\odot}) \le -0.3$ and the lowest chi-square fit resulted
with for a value of $log (M/M_{\odot})=$ -0.25 ($M$ = 0.56 $M_{\odot}$). 
The best estimate of the power-law slopes are $\alpha = -2.41 \pm$0.50 (dotted line)
for higher masses, and $\alpha = -1.25 \pm$0.19 (dashed) for lower masses, 
where $dN/dM$ = $C  M^{\alpha}$.

We also fitted the Galactic bulge IMF by using a log-normal function described by the functional form:
\begin{equation}
\xi (log M) = C \exp \left \{ -\frac{[log (M) - log(M_c)]^{2}}{2 \sigma ^{2}} \right \}
\end{equation}
with $M_c$ = 0.25$\pm$0.07 and $\sigma$ = 0.50$\pm$0.01 (solid green line in Fig.~7).

The power-law slope for the high-mass range ($M >$ 0.56 $M_{\odot}$) agrees very well with the 
Salpeter IMF ($\alpha = -2.35$) derived for solar neighborhood stars in the mass range 
0.3 -- 10 $M_{\odot}$ (blue dashed-dotted line in Fig.~7).
The log-normal mass function derived for disk stars closer than 8 pc in the mass range 0.08 to 1.0 $M_{\odot}$ by
\citet{chabrier03, chabrier05} has $M_c$ = 0.20 and $\sigma$ = 0.55 
(dashed-double dotted red line) and agrees well with our Galactic bulge IMF. 

We derived the IMFs using the same method described in the previous section 
for different values of binary fraction, assuming 
a flat distribution of mass ratios, and a distribution given by DM.
Fig.~ 8 shows one realization for each of the IMFs derived for different binary fractions 
and the DM mass-ratio distribution.
In general, the IMF has two distinct slopes in the low- and high-mass ranges, and the slopes
have only a weak dependence on the assumed mass-ratio distribution for the binaries. 
If we assume the DM mass-ratio distribution for the binary components, the slopes of the IMF 
at higher masses are $-2.25, -2.36, -2.41$ and $-2.53$, for a bulge binary fraction of 0, 
30,  50 and 100\%, respectively. If we assume a flat mass-ratio distribution, the slopes 
change only by 1 to 4\% for binary mass fractions of 0 to 100\%. The corresponding slopes in the 
low-mass range are $-0.89,  -1.12, -1.25$ and $-1.51$ for the DM mass-ratio distribution, 
and they change by $3-4\%$ for a flat mass-ratio distribution. Full details including the error bars are given in Table 1.
These results indicate that the effect of the presence of unresolved binaries 
is more pronounced in the low-mass range ($\sim$50\%), than in the high-mass range ($\sim$12\%).
In the rest of the discussion, we use the IMF derived by assuming a binary fraction of  50\%
and a DM mass-ratio distribution. As discussed in \S 5.1, the presence of a substantial
fraction of He-core WDs in the Galactic bulge suggests that the
fraction of binaries in the bulge is larger than 30\%.

\section{Comparison with other Initial mass functions}
\subsection{Galactic bulge}

The Galactic bulge mass function was first measured by HO98 based on a set of observations of the 
Baade's window $(l = 1\deg,b = -4\deg)$ collected with the Wide Field Planetary Camera 2 (WFPC2) on board HST. 
These data allowed them to derive a luminosity function down to $F814W \sim$ 24.3, corresponding to $M \sim$ 0.3 $M_{\odot}$.  No information on proper motions was available, so disk stars are included in their study.
But they applied a correction for the presence of unresolved binaries 
and found that the IMF of the bulge has a power-law slope of $\alpha = -2.2$ in the high-mass range.
The slope of the IMF flattens at $\sim 0.7 M_{\odot}$, with
$\alpha = -0.9$ for a fraction of binaries of 0\% and $-1.3$ for 50\% (see Table~2). 
HO98 result for an assumed fraction of binaries of 50\% agrees quite well, 
within the uncertainties, with what we obtained in our analysis for the same
assumption on binaries, but the changing of power-law slope occurs at lower masses, 
$\sim$ 0.56 $M_{\odot}$, in our bulge IMF.

A second study on the Galactic bulge mass function was published by ZO00, based on a set of observations collected 
in the $F110W$ and $F160W$ filters with NICMOS on board HST, covering a 22.5" $\times$ 22.5"
field of view in a region of the bulge South of the Baade's window $(l = 0.277\deg, b = -6.167\deg)$.
To convert magnitudes to masses they used a mass-luminosity relation based on the same 
stellar models adopted in this investigation. They also did not have propel-motion
information to separate bulge from disk stars, nor did they apply a correction for the presence of unresolved binaries.
However, ZO00 applied an overall reduction of the luminosity function by 11\% for magnitudes
brighter than $J < $ 17, to take into account the contamination by disk stars.
By fitting the IMF with a single power law they obtained a slope of $\alpha = -1.33\pm$0.07, 
over the mass range 0.15 $< M/M_{\odot} <$ 1.0, while by using two different power laws they
obtained $\alpha = -2.00\pm$0.23 for masses $M >$ 0.5 $M_{\odot}$, and 
$\alpha = -1.43\pm$0.13 for lower masses (see Table~2). 
If we fit our IMF by using a single power law for the entire mass range (0.15 $< M/M_{\odot} <$ 1.0), 
we obtain a range of slopes from $\alpha = -1.14\pm$0.10 for no binaries to  $\alpha = -1.56\pm$0.10 for 100\% of binaries.
The slope of the IMF not corrected for the presence of unresolved binaries is then shallower compared to the slope of the
IMF obtained by ZO00 ($-1.14$ vs $-1.33$). 
Moreover, the same IMF shows a much shallower slope in the low-mass regime
compared to ZO00 mass function ($-0.89$ vs $-1.43$). 
{This discrepancy might be due to the residual contamination by disk stars of ZO00 sample.
Part of the difference could also be due to an intrinsic difference of stars observed by ZO00 and 
stars in the SWEEPS field. From spectra collected by our group the stars in this region of the bulge have a similar 
metallicity distribution as the stars in the Baade's Window, with main peaks at 
$[M/H] \approx -0.4, 0$ and 0.3 \citep{hill11, bensby13, ness13}. 
The metallicity distribution of the region of the Galactic bulge observed by ZO00 
shows a decrease in the fraction of metal-rich stars, with the average metallicity decreasing 
from $[Fe/H] \sim$ $+$0.03 in the Baade's window down to $[Fe/H] \sim$ $-$0.12 \citep{zoccali08}.
However, such a small difference in the metallicity distribution cannot account for a
$\sim$ 20\% difference in the IMF slope.

\subsection{Galactic disk}

The Galactic disk mass function has been constrained in the low-mass regime 
down to the hydrogen-burning limit and in the brown dwarf regime by various studies. \citet{salpeter55} derived the 
``original mass function" for solar neighborhood stars in the range 0.3 $\lesssim M/M_{\odot} \lesssim$ 10 
and fitted it by using a single power law with a slope of $\alpha = -2.35$. Later studies found that the 
disk mass function can be reproduced either by a segmented power law or by a log-normal function.
Table~2 lists the power law slopes and the characteristic masses and sigmas used by 
different studies to fit the Galactic disk mass function.
\citet{kroupa93} and later \citet{kroupa01} derived the IMF for disk stars within 5.2 pc of the Sun by 
taking into account a correction for the presence of unresolved binaries and fitting it with a power law
with a slope of $\alpha = -2.2\pm$0.3 in the mass range 0.5 $< M/M_{\odot} <$ 1.0 and of
$\alpha =  -1.3\pm$0.5 in the range 0.08 $<M/M_{\odot}<$ 0.5.
\citet{gould97} based their study of the Galactic disk mass function on photometry collected with the 
WFPC2 and WFPC1 on board  HST.
They observed a sample of 337 stars distributed in different regions of the disk and 
found a mass function with a slope close to Salpeter, $\alpha \sim -2.2$, for masses
in the range 0.6 $< M/M_{\odot} <$1.0, and $\alpha \sim -0.9$ for lower masses.
 \citet{reid02} observed a sample of 558 main-sequence stars in the solar neighborhood 
in the mass range 0.1 $< M/M_{\odot} <$ 3.0 and found that a power law with a slope
of $\alpha =  -1.3$ fits the mass function in the low-mass range, i.e. for 
stars with M $<$ 0.7  $M_{\odot}$.
 \citet{chabrier05} adopted a log-normal function to fit the Galactic disk IMF 
for single stars in the mass range 0.08  $< M/M_{\odot} <$ 1.0, and found a characteristic mass 
$M_c =$ 0.20$\pm$0.02, and $\sigma =$0.55$\pm$0.05.

More recent analyses based on the Sloan Digital Sky Survey (SDSS) and the Two micron all sky survey
(2MASS) data confirmed previous results, showing that the Galactic disk mass function can be 
reproduced either by assuming a segmented power law with slopes of $\alpha = -2.04/-2.66$ and
$\alpha =  -0.8/-0.98$, for the high- and low-mass range, respectively, or by a log-normal function 
with $M_c =$  0.20/0.50, $\sigma =$  0.22/0.37 (\citealt{covey08, bochanski10}).

The IMF we derived for the Galactic bulge is in very good agreement, within uncertainties, 
with the mass function obtained by \citet{kroupa01}  and  \citet{chabrier05} for the disk.
On the other hand, the mass functions derived for the disk by \citet{covey08} and \citet{bochanski10} 
have a slightly shallower slope compared to our bulge IMF in the low-mass regime (see Table~2), 
although the two mass functions would agree at lower masses by assuming the  presence of no binaries in the bulge.

\section{The stellar mass--to--light ratio of the Galactic bulge}

The stellar mass--to--light ratio ($M/L$) is an important parameter of a stellar population and depends 
on its IMF. We used the mass function derived in this work and the total luminosity of stars observed 
in the SWEEPS field to estimate the stellar $M/L$ of the Galactic bulge in the $F814W$ and the $F606W$ filters.
We obtain a total mass for bulge stars in the SWEEPS field included in the mass range adopted to estimate the IMF, 
0.16 $\le M/M_{\odot} \le$ 1.0, of 137,527$\pm$23,400 $M_{\odot}$. By extrapolating the IMF with a power-law slope of $\alpha =-1.25$
down to the hydrogen burning limit, we get an extra mass of 14,310$\pm$2,400 $M_{\odot}$, for a total mass of $\approx$ 
152,000$\pm$23,500 $M_{\odot}$ included in the 0.10 $\le M/M_{\odot} \le $ 1.0 mass range. 
Uncertainties take into account the error budget of the derived IMF. 
A constant mass of 1.0 $M_{\odot}$ is assumed for bulge sub- and red-giant stars and red 
clump stars, for a total mass of 4,116 $M_{\odot}$. We do not take into account the mass loss along the RGB, 
but since the total mass of the giants is already very small compared to the mass of the MS stars, this has no
effect on the final derivation of the mass-to-light ratio.
We then assume that the IMF of the Galactic bulge has a constant Salpeter power-law slope for 
masses larger than 1.0 $M_{\odot}$ and up to 120 $M_{\odot}$,
and we integrate the IMF to obtain the number of stars that formed in this mass range. 
To estimate the mass currently in stellar remnants in the bulge we follow the prescriptions of \citet{percival09}:
stars with mass (i) 1 $< M/M_{\odot} \le$ 10, the remnant is a white dwarf; 
(ii) 10 $< M/M_{\odot} \le$ 25,  the remnant is a neutron star, and  (iii) $M >$ 25 $M_{\odot}$,
the remnant is a black hole. 
In order to estimate the mass of white dwarf remnants, we used the  
initial--to--final mass relation by \citet{salaris09}, $M_f =$ 0.084 $M_i$ + 0.466 for initial masses
less than 7 $M_{\odot}$ and a constant final mass of 1.3 $M_{\odot}$  for initial masses in the 
range 7 $< M/M_{\odot} \le $ 10, obtaining a total mass in white dwarfs of 53,912$\pm$9,200 $M_{\odot}$.
For neutron stars we assumed a constant mass of 1.4 $M_{\odot}$  and for black holes a mass
equal to 1/3 of the initial mass, obtaining total remnant 
masses of 3,905$\pm$600 and 11,151$\pm$1,900 $M_{\odot}$ for neutron stars and
black holes, respectively.
By using the aforementioned values we found that the total stellar mass in the 
bulge SWEEPS field is $M =$ 228,814 $\pm$25,300 $M _{\odot}$.

\begin{table}
\begin{center}
\caption{Power-law slopes of the IMFs derived by assuming different binary fractions and 
mass-ratio distributions for the Galactic bulge.}           % title of Table
\label{table:1}      % is used to refer this table in the text
\begin{tabular}{l c c c}        % centered columns (12 columns)
\hline\hline                 % inserts double horizontal lines
Binary fraction  &   Mass-ratio & $\alpha_{High}$  &  $\alpha_{Low}$    \\    % table heading 
\hline\hline
0                      &     \ldots     &    $-2.25\pm$0.50  & $-0.89\pm$0.20  \\
30                    &      DM       &    $-2.36\pm$0.51  & $-1.12\pm$0.19  \\
50                    &      DM       &    $-2.41\pm$0.50  & $-1.25\pm$0.19  \\
100                  &      DM       &    $-2.53\pm$0.51  & $-1.51\pm$0.20  \\
30                    &      Flat       &    $-2.39\pm$0.51  & $-1.16\pm$0.19  \\
50                    &      Flat       &    $-2.45\pm$0.51  & $-1.29\pm$0.19  \\
100                  &      Flat       &    $-2.62\pm$0.52  & $-1.55\pm$0.20  \\
\hline\hline                 % inserts double horizontal lines
\end{tabular}
\end{center}
\end{table}

We estimated the flux emitted by bulge stars in the SWEEPS field by using the proper-motion
cleaned photometric catalog corrected for the total fraction of stars and for completeness.
We thus obtained a total luminosity of $L_{F814W} \approx$ 104,000$\pm$2,000 $L_{\odot}$  
and $L_{F606W} \approx$ 71,000$\pm$1,400 $L_{\odot}$.

The stellar mass--to--light values based on our IMF and the photometric catalog for the SWEEPS
field are then $M/L_{F814W} =$ 2.2$\pm$0.3 and $M/L_{F606W} =$ 3.2$\pm$0.5.

We estimated the stellar mass included in our field by also using two other assumptions for 
the mass distribution at masses larger than 1 $M_{\odot}$: constant slopes of $\alpha =$-2.0 and 
of $\alpha =$-2.7. In the first case, we obtain a larger total stellar mass, $M =$ 254,505 $\pm$28,100, 
and larger mass--to--light values, $M/L_{F814W} =$ 2.4$\pm$0.4 and $M/L_{F606W}$ = 3.6$\pm$0.6, 
while in the second case we obtain smaller values, 
$M =$ 219,079 $\pm$24,200, $M/L_{F814W} =$ 2.1$\pm$0.3 and $M/L_{F606W}$ = 3.1$\pm$0.5. 
The total mass of the observed field and the stellar mass--to--light values estimated 
for the different cases are listed in Table~3.

Finally, we also explored a more theoretical route and we computed the average 
bulge luminosity in the SWEEPS field by using two different synthetic population codes by 
\citet{cignoni13} and BASTI.
For both simulations we generated a fake stellar population with properties resembling those in the Galactic bulge: 
solar metallicity, constant star formation rate between 12 and 10 Gyr, our IMF, 
a binary fraction of 50\%, distance modulus of 14.45 and reddening $E(B-V)=$ 0.5. 

In the first case we used the latest PARSEC stellar models \citep{bressan12}. 
Simulations were run until the number of stars in the magnitude range 
20 $\le F606W \le$ 22 matched the observed number ($\sim$ 25000 stars). This experiment was repeated 
1,000 times. We found average values of $L_{F606W} \sim 58,900$ and $L_{F814W} 
 \sim 92,800$. 
In order to evaluate the effect of metallicity dispersion we also tested different $Z$ values, namely 0.008 and 0.03, 
corresponding to the metal-poor and metal-rich peaks of the metallicity distribution of the considered field.
In these cases we found  $L_{F606W} \sim70,800 $ and $L_{F814W}  
\sim  101,600$ for the former metallicity, and $ L_{F606W}  \sim  53,800 $ and $ L_{F814W} \sim  87,500$ for the latter.  
As expected, lowering the metallicity causes an increase in the luminosity of the system.
Interestingly enough, luminosity values estimated for the lower metallicity, $Z = 0.008$, 
agree quite well with the observed values, while values for the higher metallicities are systematically lower 
than our flux estimates. 
A part of this discrepancy may be due to the possibility that a few very bright thin-disk stars are still
contaminating our data, raising the inferred luminosities. 
In addition, the actual PARSEC models miss 
the asymptotic giant branch stellar phase, hence the predicted luminosities are likely to be underestimated.

We repeated the same experiment using the BASTI models for the three different
metallicites, 
and obtained  $L_{F606W}$  and  $L_{F814W}$  values as
$\sim 74,500$ and $\sim 103,000$ for Z=0.008, 
$\sim 62,500$ and $ \sim  89,700$ for Z=0.02, and 
$ \sim  57,200$ and $ \sim$ 86,200 for Z=0.03. In this case the luminosity estimates for the lower metallicity are
also in very good agreement with the observed values, while the luminosities obtained for the 
solar and higher metallicites are systematically lower. On the other hand, the luminosity estimates for 
the three metallicites based on the two different sets of models agree very well.

Summarizing, we found a stellar mass--to--light ratios included in the range 2.1$< M/L_{F814W} <$ 2.4 and
3.1$< M/L_{F606W} <$ 3.6 according to different assumption on the slope of  the IMF for masses larger than 1$M_{\odot}$.
These are likely to be slightly lower estimates of the real stellar mass--light budget of  bulge since a few bright disk stars 
might still be contaminating our luminosity estimate.
These values agree quite well, within the uncertainties, with the estimates provided by ZO00 
in the Johnson $V$ filter, $M/L_V \sim$ 3.4, by using their IMF with a single 
slope of $\alpha =$-1.33 below 1 $M_{\odot}$, and by assuming a constant 
Salpeter IMF for stars more massive than 1 $M_{\odot}$.

\begin{table*}
\begin{center}
\caption{List of the different mass functions derived for the Galactic bulge and disk}           % title of Table
\label{table:2}      % is used to refer this table in the text
\begin{tabular}{l c c c c c c c}        % centered columns (12 columns)
\hline\hline                 % inserts double horizontal lines
  Reference  &   Mass range & $\alpha_{High}$  &  $\alpha_{Low}$  & $M_{break}$   & $\alpha$ & $M_c$ & $\sigma$  \\    % table heading 
%                     &                              &                             &    $M_{\odot}$  &               \\  
\hline\hline
\multicolumn{8}{c}{Galactic bulge} \\
\hline
This work                        & $0.15 - 1.0$  & $-2.41\pm$0.50  & $-1.25\pm$0.19 & 0.56 & \ldots &  0.25$\pm$0.07 & 0.50$\pm$0.01  \\
Holtzman et al. (1998)  & $0.30 - 1.0$  &  $- 2.2$ & $-1.3$   & 0.7     & \ldots & \ldots & \ldots \\
 Zoccali et al. (2000)     &  0.15 - 1.0      &  \dots  &  \ldots  &  \dots  &  $-1.33\pm$0.07 &  \ldots & \ldots  \\
\hline
\multicolumn{8}{c}{Galactic disk} \\
\hline
Salpeter (1955)                    & 0.30 - 10 &  \ldots  &  \ldots  &  \ldots  & $-2.35$ & \ldots  &  \ldots   \\
Kroupa et al. (1993, 2001)  & 0.08 - 1.0  & $-2.3\pm$0.3   & $-1.3\pm$0.5   & 0.5     & \ldots & \ldots \\
Gould et al. (1997)               & 0.08 - 1.0 &  $-2.2$  & $-0.9$   & 0.6     & \ldots & \ldots \\
Reid et al. (2002)           & 0.10 - 3.0 & \ldots  &  \ldots    &  \ldots  &  $-1.3$ & \ldots & \ldots \\
Chabrier (2005)             & 0.10 - 1.0 & \ldots  &  \ldots  &  \ldots  & \ldots  & 0.20$\pm$0.02 & 0.55$\pm$0.05 \\
Covey et al. (2008)        & 0.10 - 0.7 & $-2.04$   &  $-0.8$  & 0.32  &  $-1.1$  &  0.20 - 0.50 & 0.22 - 0.37 \\
Bochanski et al. (2010) & 0.10 - 0.8 & $-2.66\pm$0.10 & $-0.98\pm$0.10 & 0.32 & \ldots & 0.18$\pm$0.02 & 0.34$\pm$0.05 \\
\hline\hline                 % inserts double horizontal lines
\end{tabular}
\end{center}
\end{table*}

\section{Microlensing optical depth}

Several thousand microlensing events have been detected to date towards the
Galactic bulge, mainly by the OGLE \citep{udalski15} and MOA collaborations \citep{bond01,sako08,sumi13}. 
These microlensing
events have been used by several investigators to derive the total mass 
budget as well as the mass function of the lenses.

\citet{paczynski94}, based on a small number of 
microlensing events, noticed that the observed microlensing optical depth is in
excess of the theoretical estimates, indicating a much higher efficiency for
microlensing by either bulge or disk lenses.  This issue has been further
investigated in recent years by several groups \citep{Wyrzykowski15, sumi13}.
 A helpful hint comes from the latest study by Wyrzykowski et al., 
 which shows a dependence of the mean microlensing timescale on the
Galactic latitude. This signals an increasing contribution from disk lenses closer
to the plane relative to the height of the disk, which needs to be taken into
account in the estimation of timescales and optical depths. 

Since the timescale of the microlensing event is proportional to the square root
of the mass of the lens, the timescales can be used for a statistical estimate of
the mass function of the lenses. \citet{zhao95} and \citet{hangould96} 
used this approach and reported a mass function with
a slope of $-2.0$ and a cutoff at $\sim 0.1 M_\odot$.  \citet{calchi08} 
also attempted to fit the observed timescales of the microlensing
events with a power-law distribution of lens masses and obtained a slope of $-1.7$
for the distribution. As pointed out by ZO00, there may be an
extra bias in the observed timescales due to blending in the ground-based
observations, which causes the times scales to appear shorter than they actually
are.  This leads to an underestimation of the lens masses. 
Even so, the derived slope from microlensing observations is in between the 
two slopes of $\alpha =$ -2.41 and $ -1.25$ derived here, and thus seems 
consistent. It would be interesting to extend this microlensing analysis 
to the currently available list of all the observed microlensing events. 

Finally, we note that the microlensing optical depth comes not only from the
living main-sequence stars, but also from the white dwarfs, neutron stars and
black holes. The mass-to-light ratio derived in this paper should help in deriving
a more correct estimate of the microlensing optical depth.

\begin{table}
\begin{center}
\caption{Stellar mass estimates and mass--to--light values for the Galactic bulge for
different assumed IMF slopes for $M >$ 1 $M_{\odot}$.}           % title of Table
\label{table:3}      % is used to refer this table in the text
\begin{tabular}{l c c c}        % centered columns (12 columns)
\hline\hline                 % inserts double horizontal lines
$\alpha$  &   Stellar mass & $M/L_{F814W}$  &  $M/L_{F606W}$    \\    % table heading 
\hline\hline
Salpeter    &  228,814 $\pm$25,300  &  2.2$\pm$0.3  &   3.2$\pm$0.5 \\
-2.0            &  254,505 $\pm$28,100  &  2.4$\pm$0.4  &   3.6$\pm$0.6 \\
-2.7            &  219,079 $\pm$24,200  &  2.1$\pm$0.3  &   3.1$\pm$0.5 \\  
\hline\hline                 % inserts double horizontal lines
\end{tabular}
\end{center}
\end{table}

\section{Discussion and conclusions}\label{concl}
We have derived the IMF of the pure bulge component down to 0.15 $M_\odot$.
The Galactic bulge IMF can be fitted by two power laws, one with a steeper slope
$\alpha = -2.41\pm$0.50 for $M \ge$ 0.56 $M_\odot$, and another with a shallower 
slope $\alpha = -1.25\pm$0.19 for the  lower masses.
A log-normal function fits the IMF too, with a characteristic mass
of $M_c =$ 0.25$\pm$0.07 and $\sigma =$ 0.50$\pm$0.01. 

The slope of the IMF at high masses is mildly affected by the assumption 
on the fraction of unresolved binaries in the bulge or the distribution of their mass ratios.
The high-mass slope ranges from $\alpha = -2.25\pm0.50$ for no binaries to $\alpha = -2.62\pm0.52$ for 
100\% of binaries in the bulge.
On the other hand, the slope at lower masses changes significantly, 
ranging from $\alpha =-0.89\pm0.20$  for no binaries to $\alpha = -1.55\pm0.20$ for 100\% of binaries.

As we noted earlier, the slope of the IMF at the very low-mass range is crucial in estimating
the mass budget of 
the Galactic bulge which contains $\approx$20\% of the mass of the Galaxy. Our deep HST observations obtained over a timescale of $\sim$9 years allowed
us to derive the mass function of the pure bulge component even in this low-mass range, which was
previously not possible.

The shape of the Galactic bulge IMF we derived in this work is in good agreement, 
within the uncertainties, with the IMFs derived previously by HO98 for the Baade's window, but
our mass function extends to lower masses
and it is purely based on bulge members with negligible contamination from disk stars. 
On the other hand, our IMF not corrected for the presence of unresolved binaries 
shows a slightly shallower slope compared to ZO00 IMF ($-1.14$ vs $ -1.33$).
This difference could be due to a small residual contamination by disk stars of the ZO00 sample, or
to some intrinsic differences in the stars in the field observed by ZO00 and stars in the SWEEPS field.

Our bulge IMF is in very good agreement with the  mass function derived for the Galactic disk 
by \citet{kroupa01} and \citet{chabrier03} in the entire mass range, while it is steeper in the very low-mass regime compared to the mass functions
derived for the disk by \citet{gould97} and \citet{reid02}. 
The PDMFs derived in the more recent studies of \citet{covey08} and \citet{bochanski10} agree quite well with our 
IMF for the Galactic bulge in the high-mass range, but they still show a shallower slope in the low mass range.

The characterization of the IMF in different stellar environments is fundamental for investigating
if the IMF has a dependence on the stellar metallicity and/or age. 
The recent work of \citet{kalirai13} showed that the IMF of the Small Magellanic Cloud (SMC, $-1.5\lesssim [Fe/H] \gtrsim -1.0$)
is shallower than the Salpeter mass function,  $\alpha = -1.9$, down to  $\approx 0.4$ $M_{\odot}$, and
does not show evidence for a turn-over in the very low-mass regime.
Furthermore, \citet{geha13} showed that the IMF of two metal-poor ($[Fe/H] < -2.0$) ultra faint galaxies, Hercules and Leo IV,  
are even shallower, having a slope in the range $\alpha = -1.2$ to $1.3$ for masses larger than $\approx 0.5$ $M_{\odot}$. 
In the higher-mass range (M $> 0.5 M_{\odot}$) where the mass function of these galaxies is well measured,
our bulge IMF is steeper than both the IMFs of the intermediate-metallicity environment of the SMC ($-2.41$ vs $-1.9$) 
and the metal-poor environment of the ultra-faint galaxies ($-2.41$ vs $-1.3$ to $-1.2$), pointing towards a 
variation of the IMF with the global average metallicity of the stellar population. However, more 
data are needed to sample the IMF down to lower masses, i.e. 0.1 $M_{\odot}$, in the different environments, 
to confirm this preliminary result.

We then used the derived IMF to estimate the  stellar mass--to--light ratios of the Galactic bulge. 
We obtained for the two filters, values included in the range 2.1 $\le M/L_{F814W} \le$ 2.4 and
3.1 $\le M/L_{F606W} \le$ 3.6 according to different assumption on the slope of  the IMF for masses larger than 1 $M_{\odot}$.

The shape of the mass function derived from microlensing observations has
large uncertainties but is in consistent with the observed bulge IMF presented here.

\acknowledgments
This study was supported by NASA through grants GO-9750 and GO-12586 from the Space
Telescope Science Institute, which is operated by AURA, Inc., under NASA contract NAS~5-26555.
SC and RB thank financial support from PRIN-INAF2014 (PI: S. Cassisi).
We thank the anonymous referee for helpful suggestions which led to an improved version of the paper.

%%%%%%%%%%%%%%
% References %
%%%%%%%%%%%%%%

\clearpage
\bibliographystyle{aa}

\bibliography{calamida}

\end{document}